\documentclass[english,twocolumn,prl,showpacs]{revtex4}
\usepackage[latin1]{inputenc}
\usepackage{amsmath}
\usepackage{graphicx}

\batchmode

\usepackage{babel}
\makeatother

\begin{document}

\title{Effective Quantum Dynamics of two Brownian particles}

\author{O. S. Duarte and A. O. Caldeira}

\affiliation{Departamento de F\'{\i}sica da Matéria Condensada, Instituto de F\'{\i}sica
Gleb Wataghin, Universidade Estadual de Campinas, CEP 13083-970, Campinas-SP,
Brazil}

\begin{abstract}
We use the system-plus-reservoir approach to study the quantum
dynamics of a bipartite continuous variable system (two generic
particles). We present an extension of the traditional model of a
bath of oscillators which is capable of inducing an effective
coupling between the two parts of the system depending on the
choice made for the spectral density of the bath. The coupling is
nonlinear in the system variables and an exponential dependence on
these variables is imposed in order to guarantee the translational
invariance of the model if the two particles are not subject to
any external potential. The reduced density operator is obtained
by the  functional integral method. The dynamical susceptibility
of the reservoir is modelled in order to introduce, besides a
characteristic frequency, a characteristic length that determines
if the effective interaction potential is strong enough to induce
entanglement between the particles. Our model provides a criterion
of distance for identifying in which cases a common environment
can induce entanglement. Three regimes are found: the short
distance regime, equivalent to a bilinear system-reservoir
coupling, the long distance regime in which the particles act like
coupled to independent reservoirs and the intermediate regime
suitable for the competition between decoherence and
induced-entanglement.
\end{abstract}

\pacs{05.40.Jc, 03.67.Bg, 03.65.Yz}

\maketitle

\section{Introduction}

The usual model of Brownian motion has been successfully used to
describe general properties, classical or quantum mechanical, of
dissipative systems with only one degree of freedom subject to
arbitrary potentials \cite{annals,physica,Leggett,Fisher,Garg}.
Indeed, it has been extensively shown in the literature that,
within the range of interest, other approaches to dealing with
dissipative systems described by a single dynamical variable
always furnish us with the same results as those obtained by the
bath of oscillators with a properly chosen spectral function
\cite{solitons,Hedegard,Guinea,Weiss}. However, in a previous
paper \cite{Duarte} was shown that the model of independent
oscillators coupled bilinearly in coordinates to the system of
interest is inappropriate to dealing with dissipative system in
which two independent degrees of freedom are considered. There, a
generalization of the usual model was developed extending the
coupling to be nonlinear in the system variables and modifying the
spectral function to mimic the low frequency limit of the response
function of an interacting oscillators bath. The importance of
understanding the dynamics of dissipative bipartite systems
arises, for example, from quantum information and computation
where it is mandatory to study the interplay between the
decoherence and entanglement induced by the environment.

It is our intention in this paper to investigate the quantum
behavior of a system composed by two independent particles
immersed in a common environment. We use the path integral
approach to find the reduced density operator, in coordinate
representation, of the two particle system. The propagator that
controls the temporal evolution of the system is written, as
usual, in terms of the influence functional that contains the
effects of the bath. In order to guarantee that each particle
behave as a Brownian particle, if the other is absent, the
influence functional can be expressed as a triple product where
two factors are equivalent to the result one would have obtained
by coupling each particle separately to a bath of noninteracting
harmonic oscillators in the usual way and the third factor
introduces and effective interaction between the particles
mediated by the reservoir. The latter behaves as a source of
quantum correlations and allows us to find, for example, entangled
states for systems initially prepared in a separable state.

As an application we calculate the temporal evolution of the entanglement
(logarithmic negativity) for a Gaussian two mode state. The results
show a competition between decoherence and induced entanglement that
mainly depends on the temperature and the average distance between
the particles.

\section{model and exact density operator}

In a previous paper \cite{Duarte}, we presented a generalized
system-reservoir model suitable for the study of two uncoupled
particles interacting with a common bath. There, it was
demonstrated that the traditional bilinear coupling is not
appropriate to handle the effective interaction mediated by the
bath and the solution was to introduce a nonlinear coupling in the
system variables. The principal features of the model will be
reviewed in this section.

\subsection{Model}

The Hamiltonian for the complete system is given by

\begin{equation}
H_{S}+H_{R}+H_{I}.\label{Htotal}\end{equation}
 $H_{S}$ is the Hamiltonian of the system of interest, $H_{R}$ is
the Hamiltonian of the reservoir, which will be described as a
symmetrized collection of independent harmonic modes
\begin{equation}
H_{R}=\sum_{k=1}^{N}\left[\frac{p_{k}p_{-k}}{2m_{k}}+\frac{1}{2}m_{k}\omega_{k}^{2}R_{k}R_{-k}\right],\label{HR}\end{equation}
and $H_{I}$ is the interaction Hamiltonian which can be written in
two equivalent forms \cite{annals,physica}; a coordinate-velocity
coupling or a coordinate-coordinate coupling plus a new quadratic
term which is necessary to preserve the translational invariance
of (\ref{Htotal}) when the system of interest is not acted by any
external force. The two forms are linked by a simple canonical
transformation ($P_{k}\rightarrow m_{k}\omega_{k}R_{k}$ and
$R_{k}\rightarrow\frac{P_{k}}{m_{k}\omega_{k}}$).

In our generalization of the previous model, the system of interest
with a single degree of freedom will be represented by the free particle
Hamiltonian \begin{eqnarray}
H_{S} & = & \frac{P^{2}}{2M}.\label{Hs}\end{eqnarray}

For the coupling term we assume the interaction Hamiltonian  \begin{align}
&H_{I}=\frac{1}{2}\sum_{k=1}^{N}\left(C_{-k}(x)R_{k}+C_{k}(x)R_{-k}\right)\nonumber\\
&+\sum_{k=1}^{N}\frac{C_{k}(x)C_{-k}(x)}{2m_{k}\omega_{k}^{2}}.\label{HI-cc}\end{align}
\emph{ }In order to represent the effect of a local interaction of
the particle with a spatially homogeneous environment we choose
\begin{eqnarray} C_{k}(x) & = &
\kappa_{k}e^{ikx}.\label{acoplamento}\end{eqnarray} With this
choice it is straightforward to show that the entire system is
translationally invariant when there is not an external potential
\cite{Duarte}. As a consequence of (\ref{acoplamento}) the
potential renormalization in (\ref{HI-cc}) is a constant and
therefore does not contribute to the particle dynamics. Actually,
this reflects the translational invariance of whole system. A
coupling like (\ref{acoplamento}) appears, for example, when one
deals with the interaction of a particle with the density operator
of a fermionic bath \cite{Hedegard} or in the traditional polaron
problem\cite{Weiss}.

The classical equations of motion and the implications of the
coupling (\ref{acoplamento}) were studied in \cite{Duarte} for
one and two particles. There the nonlocal influence of the bath
appears only when a second particle is present in the environment.
Here we show that the nonlocal effects are also important for a
single particle when the quantum behavior is studied but the local
form\cite{annals} is recovered in a suitable limit.

\subsection{One Particle Density Operator}

In this section we use the functional integral method developed by
Feynman and Vernon \cite{Feynman-Vernon} in order to find the reduced density operator of the
system of interest. The time evolution of the total density
operator is given by \begin{equation}
\rho(t)=\exp\left(-iHt/\hbar\right)\rho(0)\exp\left(iHt/\hbar\right),\label{rho(t)}\end{equation}
and in coordinate representation we have \begin{align}
\left\langle x,\mathbf{R}\right|\rho(t)\left|y,\mathbf{Q}\right\rangle  & =\iiiint dx'dy'd\mathbf{R}'d\mathbf{Q}'\mathbf{K}(x,\mathbf{R},t;x',\mathbf{R}',0)\nonumber \\
 & *\left\langle x',\mathbf{R}'\right|\rho(0)\left|y',\mathbf{Q}'\right\rangle \mathbf{K}^{*}(y,\mathbf{Q},t;y',\mathbf{Q}',0),\label{rho(x,t)}\end{align}
where $\mathbf{R}$ is a $N$-dimensional vector representing each
degree of freedom of the bath and $\mathbf{K}$ is the coordinate
representation of the time evolution operator. The reduced density
operator emerges when we eliminate the bath variables. In the
coordinate representation this procedure corresponds to doing
$\mathbf{R=}\mathbf{Q}$ in (\ref{rho(x,t)}) and to integrate over
all possible values of $\mathbf{R}$. The reduced density operator
at time  $t$ depends on the total density operator at $t=0$ and
for simplicity we assume a separable initial condition
\begin{equation}
\rho(0)=\tilde{\rho}(0)\rho_{R}(0),\label{rho init}\end{equation}
where $\tilde{\rho}(0)$ is the initial particle density operator
and $\rho_{R}(0)=Z_{R}^{-1}\exp(-\beta H_{R})$ is the bath density
operator, which is assumed to be in thermal equilibrium, before
the perturbation is switched on. With this assumptions we can
write\begin{equation} \tilde{\rho}(x,y,t)=\iint
dx'dy'J(x,y,t;x',y',0)\tilde{\rho}(x',y',0),\label{reduced
rho}\end{equation} where $J(x,y,t;x',y',0)$ is the
superpropagator, which controls the time evolution of the reduced
density operator and can be written as a path integral
\begin{align}
J(x&,y,t;x',y',0)=\int\limits _{x'}^{x}\mathcal{D}x(t')\int\limits _{y'}^{y}\mathcal{D}y(t')*\nonumber\\
&\exp\frac{i}{\hbar}\left\{
S_{0}\left[x(t')\right]-S_{0}\left[y(t')\right]\right\}
\mathcal{F}\left[x(t'),y(t')\right]. \label{J-1}\end{align}
$S_{0}$ is the action of the isolated particle and
$\mathcal{F}\left[x(t'),y(t')\right]$ is a functional of the
particle trajectory, which contains all the bath information
\cite{Feynman-Vernon}. The influence functional has a simple
formal expression \begin{equation}
\mathcal{F}\left[x(t'),y(t')\right]=\textrm{Tr}_{R}
\left(\rho_{R}U_{RI}^{\dagger}[y(t')]U_{RI}[x(t')]\right),\label{F-f}\end{equation}
where $U_{RI}[x(t')]$ is the unitary time evolution operator of
the reservoir subjected to the influence of the system which
evolves through the Hamiltonian $H_{RI}=H_{R}+H_{I}[x(t')]$. This
means that a given trajectory $x(t')$ of the system for $0\leq
t'\leq t$ acts as a forcing term to the environment. The time
evolution of the operator $U_{RI}[x]$ is described by the
Schrödinger equation
\begin{equation}
i\hbar\frac{d}{dt}U_{RI}(t)=H_{RI}(t)U_{RI}(t),\label{URI
evol}\end{equation} with the initial condition $U_{RI}(0)=1$ and
has the formal solution
\begin{equation}
U_{RI}(t)=Te^{-i\int_{0}^{t}dt'H_{RI}(t')/\hbar},\label{URI-f}\end{equation}
where $T$ is the time ordering operator. Switching to the
interaction picture \cite{Fetter} this result can be written as
\begin{equation}
U_{RI}(t)=e^{-iH_{R}t/\hbar}Te^{-i\int_{0}^{t}dt'\tilde{H}_{I}[x(t')]/\hbar},\label{URI
sol}\end{equation} where $\tilde{H}_{I}[x(t')]=
e^{iH_{R}t/\hbar}H_{I}[x(t')]e^{-iH_{R}t/\hbar}$. Inserting
(\ref{URI sol}) into (\ref{F-f}) we have
$\mathcal{F}\left[x(t'),y(t')\right]=\textrm{Tr}_{R}\left(\rho_{R}Te^{i\int_{0}^{t}dt'\tilde{H}_{I}[y(t')]/\hbar}Te^{-i\int_{0}^{t}dt'\tilde{H}_{I}[x(t')]/\hbar}\right).$\emph{
}

Since the reservoir have many degrees of freedom we can assume
that the particle induces a weak perturbation in the environment
and expand the chronological product to second order in
$\tilde{H}_{I}$\cite{Hedegard},\begin{align}
&Te^{-i\int_{0}^{t}dt'\tilde{H}_{I}[x(t')]/\hbar}\approx1-\frac{i}{\hbar}\int_{0}^{t}dt'\tilde{H}_{I}[x(t')]\nonumber\\
&-\frac{1}{\hbar^{2}}\int_{0}^{t}dt'\int_{0}^{t'}ds\tilde{H}_{I}[x(t')]\tilde{H}_{I}[x(s)].\label{T
expan}\end{align} With these assumptions and tracing the
reservoir variables we find

\begin{widetext}\begin{eqnarray*}
\mathcal{F}\left[x(t'),y(t')\right] & \approx & 1-\frac{1}{\hbar^{2}}\int_{0}^{t}dt'\int_{0}^{t'}ds\left(\left\langle \tilde{H}_{I}[x(t')]\tilde{H}_{I}[x(s)]\right\rangle +\left\langle \tilde{H}_{I}[y(s)]\tilde{H}_{I}[y(t')]\right\rangle \right.\\
 &  & \left.-\left\langle \tilde{H}_{I}[y(t')]\tilde{H}_{I}[x(s)]\right\rangle -\left\langle \tilde{H}_{I}[y(s)]\tilde{H}_{I}[x(t')]\right\rangle \right).\end{eqnarray*}

Once we consider only terms to second order in
$\tilde{H}_{I}[x(t')]$ the influence functional above is
equivalent to \begin{eqnarray}
\mathcal{F}\left[x(t'),y(t')\right] & = & \exp\left\{ -\frac{1}{\hbar^{2}}\int_{0}^{t}dt'\int_{0}^{t'}ds\left(\left\langle \tilde{H}_{I}[x(t')]\tilde{H}_{I}[x(s)]\right\rangle +\left\langle \tilde{H}_{I}[y(s)]\tilde{H}_{I}[y(t')]\right\rangle \right.\right.\nonumber \\
 &  & \left.\left.-\left\langle \tilde{H}_{I}[y(t')]\tilde{H}_{I}[x(s)]\right\rangle -\left\langle \tilde{H}_{I}[y(s)]\tilde{H}_{I}[x(t')]\right\rangle \right)\right\} .\label{Funcional final}\end{eqnarray}
\end{widetext} 
Now we calculate the averages in (\ref{Funcional final})
considering the coupling (\ref{acoplamento}) and invoking the reservoir
translational invariance. Since we have $\left\langle R_{k}(t')R_{k'}(s)\right\rangle =0$
unless $k'=-k$, one is allowed to write
\begin{align}
&\left\langle \tilde{H}_{I}[x(t')]\tilde{H}_{I}[x(s)]\right\rangle =\frac{1}{2}\sum_{k}\left\{ C_{-k}[x(t')]C_{k}[x(s)]\right.\nonumber\\
&\left.+C_{k}[x(t')]C_{-k}[x(s)]\right\} \left\langle R_{k}(t')R_{-k}(s)\right\rangle .\end{align} 
The expression $\left\langle R_{k}(t')R_{-k}(s)\right\rangle $ can be written in
terms of the bath linear response through the
fluctuation-dissipation theorem. We call for simplicity
$\alpha_{k}(t'-s)=\left\langle R_{k}(t')R_{-k}(s)\right\rangle $
and \begin{equation}
\alpha_{k}(t'-s)=\frac{\hbar}{\pi}\int_{-\infty}^{\infty}d\omega\textrm{Im}\tilde{\chi}_{k}(\omega)
\frac{e^{-i\omega(t'-s)}}{1-e^{-\omega\hbar\beta}},\label{alpha(t-s)}\end{equation}
where $\tilde{\chi}_{k}(\omega)$ is the dynamical susceptibility
of the reservoir. So, the averages in (\ref{Funcional
final}) have the final form \begin{widetext}
\begin{equation} \left\langle
\tilde{H}_{I}[x(t')]\tilde{H}_{I}[x(s)]\right\rangle
=\frac{1}{2}\sum_{k} \left\{
C_{-k}[x(t')]C_{k}[x(s)]+C_{k}[x(t')]C_{-k}[x(s)]\right\}
\alpha_{k}(t'-s),\label{HI 1}\end{equation} and similar ones for
the other terms. Putting all those together we find the following
nonlinear influence functional\begin{align}
 & \mathcal{F}[x(t'),y(t')]=\exp\biggl\{-\frac{1}{\hbar^{2}}\int_{0}^{t}dt'\int_{0}^{t'}ds\sum_{k}\bigl[\kappa_{k}
 \kappa_{-k}\left(\cos k[x(t')-x(s)]-\cos k[y(t')-x(s)]\right)\alpha_{k}(t'-s)\nonumber \\
 & \,\,\,\,\,\,\,\,\,\,\,\,\,\,\,\,\,\,\,\,\,\,+\kappa_{k}\kappa_{-k}\left(\cos k[y(t')-y(s)]-\cos k[y(s)-x(t')]\right)
 \alpha_{k}^{\dagger}(t'-s)\bigr]\biggr\}.\label{Func influ expl}\end{align}
\end{widetext}

The function $\alpha_{k}(t'-s)$ is related to the bath linear
response by the Fourier transform (\ref{alpha(t-s)}). At this
point it is necessary to model the bath dynamical susceptibility
$\tilde{\chi}_{k}(\omega)$, since it is not our intention to
completely describe  the microscopic details of the bath. We can
assume the imaginary part of the bath dynamical susceptibility has
the form \begin{equation}
\textrm{Im}\tilde{\chi}_{k}(\omega)=f(k)\omega\theta(\omega-\Omega).\label{Im
dyn sucept}\end{equation} Here we introduce a high frequency
cutoff $\Omega$ as the characteristic frequency of the bath. The
Markov dynamics is achieved when we take the limit
$\Omega\to\infty$,\emph{ }and the function $f(k)$ responds for the
nonlocal influence of the bath.\emph{ }This approximation is
equivalent to replace the free oscillator response function by the
low frequency limit of the damped oscillator response function
\cite{Duarte}\emph{ }and allow us to separate the characteristic
time and length scales of the reservoir.\emph{ }A functional
dependence like (\ref{Im dyn sucept}) for the dynamical response
of the bath has been employed in Refs. \cite{Hedegard,Guinea} for
fermionic environments.

The parity of $\textrm{Im}\tilde{\chi}_{k}(\omega)$ allow us to write
the correlation function $\alpha_{k}(t'-s)$ as \begin{equation}
\alpha_{k}(t'-s)=\alpha_{k}^{(R)}(t'-s)+i\alpha_{k}^{(I)}(t'-s),\label{alpha sep}\end{equation}
with the real and imaginary parts defined respectively as \begin{align}
\alpha_{k}^{(R)}&(t'-s)=\nonumber\\
&\frac{\hbar}{\pi}\int_{0}^{\infty}d\omega\textrm{Im}\tilde{\chi}_{k}(\omega)\cos\omega(t'-s)\coth(\hbar\beta\omega/2)\label{parte Re}\end{align}
 and\begin{eqnarray}
\alpha_{k}^{(I)}(t'-s) & = &
-\frac{\hbar}{\pi}\int_{0}^{\infty}d\omega\textrm{Im}\tilde{\chi}_{k}(\omega)\sin\omega(t'-s).\label{parte
Im}\end{eqnarray} The functional (\ref{Func influ expl}) can also
be written in terms of these real and imaginary parts. With
prescription (\ref{Im dyn sucept}) we can evaluate the frequency
integrals in the imaginary part (\ref{parte Im}) taking the limit
$\Omega\to\infty$ and considering that the coupling was switched
on at \emph{ }$t=0^{+}$. Following this procedure we have a
functional with a Markovian imaginary part and a explicit
nonlinear dependence, \begin{widetext}\begin{align*}
\mathcal{F}[x(t'),y(t')] & =\exp\biggl\{\frac{i}{2\hbar}\sum_{k}\kappa_{k}\kappa_{-k}f(k)k\int_{0}^{t}dt'\sin k[y(t')-x(t')]\left(\dot{x}(t')+\dot{y}(t')\right)\\
 & -\frac{1}{\hbar^{2}}\int_{0}^{t}dt'\int_{0}^{t'}ds\sum_{k}\kappa_{k}\kappa_{-k}\left[\cos k[x(t')-x(s)]-\cos k[y(t')-x(s)]\right.\\
 & \left.+\cos k[y(t')-y(s)]-\cos k[y(s)-x(t')]\right]\alpha_{k}^{(R)}(t'-s)\biggr\}.\end{align*}
\end{widetext}This functional can be approximated assuming that the
most important trajectories ($x(t'),y(t')$) are confined within a\emph{
}region small compared to a characteristic length $k_{0}^{-1}$ introduced
in (\ref{Im dyn sucept}) through the function $f(k)$. For example,
in fermionic environments this length is related to the Fermi wave
number $k_{F}$ \cite{Hedegard,Guinea}. In this approximation, we
have $k(y(t')-x(t'))\ll1$ and the functional can be written as \begin{widetext}\begin{align}
 & \mathcal{F}[x(t'),y(t')]=\exp\left\{ \frac{i\eta}{2\hbar}\int_{0}^{t}dt'(y(t')-x(t'))\left(\dot{x}(t')+\dot{y}(t')\right)\right.\nonumber \\
 & \left.-\frac{\eta}{\hbar\pi}\int_{0}^{t}dt'\int_{0}^{t'}ds\int_{0}^{\infty}d\omega\,\omega\coth\left(\frac{\hbar\beta\omega}{2}\right)(x(t')-y(t'))\cos\omega(t'-s)(x(s)-y(s))\right\} ,\label{local funct}\end{align}
\end{widetext}where we have identified $\eta=\sum_{k}k^{2}\kappa_{k}\kappa_{-k}f(k)$.
Notice that, with this modification, we obtain a relation between
the damping constant and some microscopic parameters of the
oscillator bath. The functional (\ref{local funct}) coincides with
the result obtained\emph{ }by coupling the particle of interest
bilinearly to a bath of noninteracting harmonic oscillators with
the spectral function $J(\omega)=\eta\omega$
\cite{annals,physica}.

\subsection{Two Particle Density Operator}

Now we are going to study the dynamics of a system with two degrees
of freedom immersed in a dissipative environment. In this case the
Lagrangian of the system of interest is \begin{equation}
L_{S}=\frac{1}{2}M\dot{x}_{1}^{2}+\frac{1}{2}M\dot{x}_{2}^{2},\label{LS-2P}\end{equation}
 and the coupling term \begin{align}
L_{I}&=-\frac{1}{2}\sum_{k}\left[\left(C_{-k}(x_{1})+C_{-k}(x_{2})\right)R_{k}\right.\nonumber\\
&\left.+\left(C_{k}(x_{1})+C_{k}(x_{2})\right)R_{-k}\right].\label{LI-2P}\end{align}

Notice that we have not included any counter-term in (\ref{LI-2P})
since our system is manifestly translationally invariant. The time
evolution of the reduced density operator for the two particle
system is given by \begin{equation}
\tilde{\rho}(\mathbf{x},\mathbf{y},t)=\iint
d²xd²yJ(\mathbf{x},\mathbf{y},t;\mathbf{x}',\mathbf{y}',0)\tilde{\rho}(\mathbf{x}',\mathbf{y}',0),\label{redu
two P}\end{equation} and the superpropagator for this case is
\begin{align}
J&(\mathbf{x},\mathbf{y},t;\mathbf{x}',\mathbf{y}',0)=\int\limits _{\mathbf{x}'}^{\mathbf{x}}\mathcal{D}\mathbf{x}(t')\int\limits _{\mathbf{y}'}^{\mathbf{y}}\mathcal{D}\mathbf{y}(t')\nonumber\\
&\exp\frac{i}{\hbar}\left\{ S_{0}\left[\mathbf{x}(t')\right]\right.\left.-S_{0}\left[\mathbf{y}(t')\right]\right\} F\left[\mathbf{x}(t'),\mathbf{y}(t')\right],\label{J two P}\end{align}
where we have defined the vectors $\mathbf{x}(t')=(x_{1}(t'),x_{2}(t'))$
and $\mathbf{y}(t')=(y_{1}(t'),y_{2}(t'))$. $S_{0}$ is the action
of the isolated two particle system and $F$ is the Feynman-Vernon
influence functional, which in the operator form can be written as
\begin{equation}
F\left[\mathbf{x}(t'),\mathbf{y}(t')\right]=\textrm{Tr}_{R}\left(\rho_{R}U_{RI}^{\dagger}[\mathbf{y}(t')]U_{RI}[\mathbf{x}(t')]\right),\label{Funct
two P}\end{equation} where $U_{RI}[\mathbf{x}(t')]$ is the unitary
time evolution operator as in (\ref{URI-f}). The procedure to
calculate the influence functional for the two particle case is
completely equivalent to the one particle case and we can directly
generalize the result, keeping only terms to second order in
$H_{I}[\mathbf{x}(t')]$, as \begin{align}
 & F\left[\mathbf{x}(t'),\mathbf{y}(t')\right]=\exp\biggl\{-\frac{1}{\hbar^{2}}\int_{0}^{t}dt'\int_{0}^{t'}ds\nonumber \\
 & \times\left(\left\langle \tilde{H}_{I}[\mathbf{x}(t')]\tilde{H}_{I}[\mathbf{x}(s)]\right\rangle +\left\langle \tilde{H}_{I}[\mathbf{y}(s)]\tilde{H}_{I}[\mathbf{y}(t')]\right\rangle \right.\\
 & -\left\langle \tilde{H}_{I}[\mathbf{y}(t')]\tilde{H}_{I}[\mathbf{x}(s)]\right\rangle \left.-\left\langle \tilde{H}_{I}[\mathbf{y}(s)]\tilde{H}_{I}[\mathbf{x}(t')]\right\rangle \right)\biggr\}.\label{Funct expl two P}\end{align}
Again, we take the average over the equilibrium density matrix of
the bath and using the fluctuation dissipation theorem and the correlation
function defined in (\ref{alpha sep}), we can write the two particle
influence functional as \begin{align}
 & F\left[\mathbf{x}(t'),\mathbf{y}(t')\right]=\nonumber \\
 & \exp\left\{ -\frac{1}{\hbar^{2}}\int_{0}^{t}dt'\int_{0}^{t'}ds\tilde{K}[\mathbf{x},\mathbf{y},t',s]\alpha_{k}^{(R)}(t'-s)\right.\nonumber \\
 & \left.-\frac{i}{\hbar^{2}}\int_{0}^{t}dt'\int_{0}^{t'}ds\tilde{K}[\mathbf{x},\mathbf{y},t',s]\alpha_{k}^{(I)}(t'-s)\right\} ,\label{two funct}\end{align}
where we have defined the nonlinear kernel \begin{align}
&\tilde{K}[\mathbf{x},\mathbf{y},t',s]=\nonumber\\
&\sum_{k}\kappa_{k}\kappa_{-k}\sum_{i,j=1}^{2}\left[\cos k[x_{i}(t')-x_{j}(s)]\right.+\cos k[y_{i}(t')-y_{j}(s)]\nonumber\\
&\left.-\cos k[y_{i}(t')-x_{j}(s)]-\cos k[x_{i}(t')-y_{j}(s)]\right].\label{Kr}\end{align}
The imaginary part of (\ref{two funct}) can be reduced using the
response function (\ref{Im dyn sucept}) in a form completely analogous
to the one particle case. The resulting imaginary part of the exponent
of the influence functional can be written, in a compact form, as

\begin{equation}
-\frac{1}{2\hbar}\int_{0}^{t}dt'\tilde{L}[\mathbf{x},\mathbf{y},t']+\frac{2\Omega}{\hbar\pi}\int_{0}^{t}dt'
V[\mathbf{x},\mathbf{y},t'],\label{Im L V}\end{equation}
 where we have defined the instantaneous kernels\begin{widetext} \begin{align}
\tilde{L}[\mathbf{x},\mathbf{y},t'] & =\sum_{k}\kappa_{k}\kappa_{-k}kf(k)
\biggl(\sum_{i,j=1}^{2}\sin k[x_{i}(t')-y_{j}(t')]\left[\dot{x}_{i}(t')+\dot{y}_{j}(t')\right]\nonumber \\
 & +\sin k[x_{1}(t')-x_{2}(t')]\left[\dot{x}_{2}(t')-\dot{x}_{1}(t')\right]-\sin k[y_{1}(t')-y_{2}(t')]
 \left[\dot{y}_{2}(t')-\dot{y}_{1}(t')\right]\biggr)\label{Lr}\end{align}
and\begin{equation}
V[\mathbf{x},\mathbf{y},t']=\sum_{k}\kappa_{k}\kappa_{-k}f(k)\left(\cos k[x_{1}(t')-x_{2}(t')]-
\cos k[y_{1}(t')-y_{2}(t')]\right).\label{V[x,y,t]}\end{equation}
\end{widetext} The well-known effects of dissipation and diffusion
are included in the kernel $\tilde{L}[\mathbf{x},\mathbf{y},t']$.
Note that this term contains direct and indirect influence of
dissipation . On the other hand, the function
$V[\mathbf{x},\mathbf{y},t']$ introduces a completely new effect
that can be interpreted as an effective interaction mediated by
the environment. The functional at this point can be written as
\begin{align}
 & F\left[\mathbf{x}(t'),\mathbf{y}(t')\right]=\nonumber \\
 & \exp\left\{ -\frac{1}{\hbar^{2}}\int_{0}^{t}dt'\int_{0}^{t'}ds\tilde{K}[\mathbf{x},
 \mathbf{y},t',s]\alpha_{k}^{(R)}(t'-s)\right.\nonumber \\
 & \left.-\frac{i}{2\hbar}\int_{0}^{t}dt'\tilde{L}[\mathbf{x},\mathbf{y},t']+\frac{i2\Omega}
 {\hbar\pi}\int_{0}^{t}dt'V[\mathbf{x},\mathbf{y},t']\right\}. \label{func inst}\end{align}
The functional above is still too complex and additional
assumptions are necessary. Since we are concerned only with the
effective terms mediated by the reservoir, we can assume that each
particle trajectory is localized within a region restricted by the
characteristic length of the reservoir. That leads to the
approximation $k[x_{i}(t')-y_{i}(s)]\ll1$ and taking only the
terms up to second order in $k[x_{i}(t')-y_{i}(s)]$ the functional
can be written as \begin{align}
 & F\left[\mathbf{x}(t'),\mathbf{y}(t')\right]=\mathcal{F}\left[x_{1}(t'),y_{1}(t')\right]\mathcal{F}
 \left[x_{2}(t'),y_{2}(t')\right]\nonumber \\
 & \exp\left\{ -\frac{1}{\hbar^{2}}\int_{0}^{t}dt'\int_{0}^{t'}ds\bar{K}[\mathbf{x},\mathbf{y},t',s]
 \alpha_{k}^{(R)}(t'-s)\right.\nonumber \\
 & \left.-\frac{i}{2\hbar}\int_{0}^{t}dt'\bar{L}[\mathbf{x},\mathbf{y},t']+\frac{i2\Omega}{\hbar\pi}
 \int_{0}^{t}dt'V[\mathbf{x},\mathbf{y},t']\right\}, \label{funct inst app}\end{align}
where $\bar{K}$ and $\bar{L}$ are defined as in (\ref{Kr}) and
(\ref{Lr}), respectively, but excluding the terms with $i=j$. The
factor $\mathcal{F}\left[x_{i}(t'),y_{i}(t')\right]$ was defined
in (\ref{local funct}) and corresponds to the influence functional
within the local approximation of the Brownian particle dynamics.
The functional above is formed by three clearly distinct parts,
the first two correspond to the direct influence of the bath over
each particle separately and the third part contains the
interaction between the particles mediated by the bath.

Towards a better understanding of the interaction factor in
(\ref{funct inst app}) it is important to study the short and long
distance limits, which are obviously defined in relation to the
characteristic length of the reservoir $k_{0}^{-1}$. For the first
we assume $k[x_{i}(t')-x_{j}(s)]\ll1$ for all particle
trajectories and for all times, which means that the particles are
very close and the dynamics is essentially local. In this
approximation the superpropagator reduces to

\begin{widetext}

\begin{align*}
 & J=\exp\left\{ \frac{-i}{\hbar}h(X,Y)\right\} \int\limits _{\mathbf{x}'}^{\mathbf{x}}
 \int\limits _{\mathbf{y}}^{\mathbf{y}}\mathcal{D}\mathbf{x}(t')\mathcal{D}\mathbf{y}(t')
 \exp\frac{i}{\hbar}\Biggl\{ S_{0}\left[\mathbf{x}(t')\right]-S_{0}\left[\mathbf{y}(t')\right]-
 \frac{\eta}{2}\int_{0}^{t}(x_{1}\dot{y}_{1}-y_{1}\dot{x}_{1})dt'-\frac{\eta}{2}
 \int_{0}^{t}(x_{2}\dot{y}_{2}-y_{2}\dot{x}_{2})dt'\\
 & \,\,\,\,\,\,\,\,\,\,\,\,\,\,\,-\frac{\eta}{2}\int_{0}^{t}dt'
 \left[x_{1}\dot{y}_{2}-y_{2}\dot{x}_{1}+x_{2}\dot{y}_{1}-y_{1}\dot{x}_{2}\right]-
 \frac{\eta\Omega}{\pi}\int_{0}^{t}dt'\left[\left(x_{1}-x_{2}\right)^{2}-\left(y_{1}-y_{2}\right)^{2}\right]\Biggr\}\\
 & \,\,\,\,\,\,\,\,\,\,\,\,\,\,\,\times\exp\left\{ -\frac{\eta}{\hbar\pi}
 \int_{0}^{t}dt'\int_{0}^{t'}ds\sum_{i,j=1}^{2}\left[x_{i}(t')-y_{i}(t')\right]K(t'-s)
 \left[x_{j}(s)-y_{j}(s)\right]\right\} ,\end{align*}
\end{widetext}where we have introduced the temperature dependent kernel
\begin{equation}
K(t'-s)=\int_{0}^{\Omega}d\omega\,\omega\coth(\hbar\beta\omega/2)\cos\omega(t'-s)\label{K(t-s)}\end{equation}
and the function \begin{align}
h(X,Y)&=\frac{\eta}{4}\left[(x_{1}+x_{2})^{2}-(x'_{1}+x'_{2})^{2}\right.\nonumber\\
&\left.-(y_{1}+y_{2})^{2}+(y'_{1}+y'_{2})^{2}\right].\label{h(X,Y)}\end{align}
Now, for simplicity, we can use two successive changes of
variables. The first is defined as
$q_{i}(t)=(x_{i}(t)+y_{i}(t))/2$,
$\,\,\xi_{i}(t)=x_{i}(t)-y_{i}(t)$ and can be interpreted as the
center and width of the wave packets respectively. The second
introduces the center of mass and relative coordinate variables
defined respectively as $r(t)=(q_{1}(t)+q_{2}(t))/2$, $\,\,
u(t)=q_{1}(t)-q_{2}(t)$ and the auxiliar variables
$\chi(t)=(\xi_{1}(t)+\xi_{2}(t))/2$, $\,\,
v(t)=\xi_{1}(t)-\xi_{2}(t)$. With these replacements the local
version of the superpropagator reads \begin{widetext}
\begin{align}
 & J=\exp\left\{ \frac{-i}{\hbar}\tilde{h}(r,\chi)\right\} \int\limits _{r'}^{r}\int\limits _{\chi'}^{\chi}\int\limits _{u'}^{u}\int\limits _{v'}^{v}\mathcal{D}r(t')\mathcal{D}\chi(t')\mathcal{D}u(t')\mathcal{D}v(t')\exp\Biggl\{-\frac{4\eta}{\hbar\pi}\int_{0}^{t}dt'\int_{0}^{t'}ds\, K(t'-s)\chi(t')\chi(s)\Biggr\}\nonumber \\
 & \exp\frac{i}{\hbar}\Biggl\{\Sigma[r,\chi,u,v]-2\eta\int_{0}^{t}dt'[\chi(t')\dot{r}(t')-r(t')\dot{\chi}(t')]-\frac{2\eta\Omega}{\pi}\int_{0}^{t}dt'u(t')v(t')dt'\Biggr\},\label{local J 2P}\end{align}

\end{widetext} with \begin{equation}
\Sigma[r,\chi,u,v]=\int_{0}^{t}M(2\dot{r}\dot{\chi}+\dot{u}\dot{v}/2)dt'\label{sigma}\end{equation}
 and\begin{equation}
\tilde{h}(r,\chi)=2\eta(r\chi-r'\chi').\label{h til}\end{equation}
From the local functional (\ref{local J 2P}) we can point out the
following features. Firstly, the stationary trajectory of $r(t)$,
deduced from the action functional in the propagator's exponent,
describes the dissipative dynamics of a particle with mass $2M$,
that is the center of mass of the two Brownian particle system.
The second, and more remarkable result is that, the relative
coordinate $u(t)$ in this approximation describes a
dissipationless dynamics of a variable acted upon by an effective
force which induces an interaction between the individual parts of
the system of interest. This effect had previously been noticed
in a classical approach and coincides with that of other works in
which the quantum evolution was studied using the master equation
approach and a bilinear system-bath coupling\cite{Horhammer,J
Paz}.

In the long distance limit we suppose that each particle is
restricted to move within separated regions. In this case the
distance $L$ between the regions is considered bigger than the
characteristic length of the reservoir and it is possible to
approximate $k[x_{i}(t')-y_{i}(s)]\gg1$. When this condition is
satisfied the interaction terms are negligible and the
superpropagator reads
\begin{widetext}\begin{align}
J= & \int\limits _{r'}^{r}\int\limits _{u'}^{u}\int\limits _{\chi'}^{\chi}\int\limits _{v'}^{v}
\mathcal{D}r(t')\mathcal{D}u(t')\mathcal{D}\chi(t')\mathcal{D}v(t')\exp\left\{ \frac{i}{\hbar}
\int_{0}^{t}dt'M\left(\frac{\dot{u}(t')\dot{v}(t')}{2}+2\dot{r}(t')\dot{\chi}(t')\right)\right\} \nonumber \\
 & \exp\left\{ -\frac{i\eta}{\hbar}\int_{0}^{t}dt'\left(\frac{\dot{u}(t')v(t')}{2}+2\dot{r}(t')\chi(t')\right)
 -\frac{\eta}{\hbar\pi}\int_{0}^{t}dt'\int_{0}^{t'}ds\left(\frac{v(s)v(t')}{2}+2\chi(s)\chi(t')\right)K(t'-s)\right\} .
 \label{J longe ru}\end{align}
\end{widetext} It is obvious that in the long distance limit the dynamics
is completely equivalent to the case of two uncoupled particles interacting
with two independent but identical environments.

Now, in order to study an approach suitable for interparticle
distances between the two  above-mentioned limits, it is necessary
to introduce new assumptions. We consider the variables
insensitive to the interparticle distance negligible in relation
to the characteristic length $k_{0}^{-1}$, i. e. the variables $v$
and $\chi$ , which are related to the width of the wave packets,
can be kept as very small when compared with $k$ and the
approximations $\cos\left[k\chi(t')\right]\approx1$,
$\sin\left[k\chi(t')\right]\approx k\chi(t')$,
$\sin\left[kv(t')/2\right]\approx kv(t')/2$ appear to be adequate
ones. On the other hand, the nonlinear terms involving the
variables $r$ and $u$ will be replaced by a phenomenological
function parametrized by the relation between the characteristic
length of the bath and the average distance between the physically
allowed regions for the particle trajectories. In this way the
superpropagator can be written as \begin{widetext} \begin{align}
 & J=\exp\left\{ \frac{i}{\hbar}\tilde{h}(r,\chi,u,v)\right\} \int\limits _{r'}^{r}\int\limits _{u'}^{u}
 \int\limits _{\chi'}^{\chi}\int\limits _{v'}^{v}\mathcal{D}r(t')\mathcal{D}u(t')\mathcal{D}\chi(t')
 \mathcal{D}v(t')\nonumber \\
 & \times\exp\frac{i}{\hbar}\left\{ M\int_{0}^{t}dt'\left(2\dot{r}\dot{\chi}+\dot{u}\dot{v}/2\right)+
 \frac{\eta}{4}\int_{0}^{t}dt'\left(4\dot{\chi}(t')r(t')+\dot{v}(t')u(t')-\dot{u}(t')v(t')-4\dot{r}(t')
 \chi(t')\right)\right.\nonumber \\
 & \,\,\,\,\left.-\frac{\eta}{2}\int_{0}^{t}dt'\left(D(k_{0}L)\dot{v}(t')u(t')+4D(k_{0}L)\dot{r}(t')
 \chi(t')\right)-\frac{2\eta\Omega}{\pi}\int_{0}^{t}dt'D(k_{0}L)u(t')v(t')\right\} \nonumber \\
 & \times\exp\left\{ -\frac{\eta}{\hbar\pi}\int_{0}^{t}dt'\int_{0}^{t'}dsK(t'-s)
 \left\{ 2\left(1+D(k_{0}L)\right)\chi(t')\chi(s)+\frac{1}{2}\left(1-D(k_{0}L)\right)v(t')v(s)\right\} \right\} .
 \label{J med}\end{align}
 \end{widetext}where we redefine \begin{align}
\tilde{h}(r,\chi,u,v)&=\frac{\eta}{4}\left(4r'\chi'-4r\chi+u'v'-uv\right)\nonumber\\
&-\frac{\eta}{2}D(k_{0}L)\left(u'v'-uv\right),\label{h til
new}\end{align} and introduce the function \begin{equation}
D(k_{0}L)=\exp\left(-k_{0}L\right),\label{D(k0L)}\end{equation}
with the parameter $L$ being a definition of the average length
between the regions where the particles move. Actually one could
use it as the initial distance between the centers of the packets
once one can make sure they will not overlap in the long run.

The functional integrals in (\ref{J med}) can be evaluated directly
since all the variables appear to second order only. The result of
this integration is \begin{widetext} \begin{align}
 & J=\mathcal{N}(t)\exp\frac{i}{\hbar}\Biggl\{\tilde{h}(r,\chi,u,v)-
 \frac{M}{\sinh\left[t\gamma_{+}/2 \right]}
 \biggl\{ \gamma_{+}\left(r'-r\right)\left(\chi e^{-\gamma_{+}t/2}-
 \chi'e^{\gamma_{+}t/2}\right)-\gamma\left(r\chi-r'\chi'\right)\sinh\left[t \gamma_{+}/2 \right]\biggr\}\nonumber \\
 & \,\,\,\,-\frac{M}{4\sinh\left[\frac{t}{2}\sqrt{\gamma_{-}^{2}-4\omega_{t}^{2}}\right]}
 \left\{ \sqrt{\gamma_{-}^{2}-4\omega_{t}^{2}}\left(u've^{-\gamma_{-}t/2}+uv'e^{\gamma_{-}t/2}\right)\right.\nonumber \\
 & \,\,\,\,-\left(\gamma_{-}-\gamma\right)\left(uv-u'v'\right)\sinh\left[\frac{t}{2}\sqrt{\gamma_{-}^{2}-
 4\omega_{t}^{2}}\right]\left.-\sqrt{\gamma_{-}^{2}-4\omega_{t}^{2}}\left(uv+u'v'\right)\cosh\left[\frac{t}{2}
 \sqrt{\gamma_{-}^{2}-4\omega_{t}^{2}}\right]\right\} \Biggr\}\nonumber \\
 & \times\exp-\frac{1}{\hbar}\biggl\{(1+D(k_{0}L))\left(A_{\chi}(t)\chi^{2}+B_{\chi}(t)\chi\chi'+C_{\chi}(t)
 \chi'^{2}\right)+(1-D(k_{0}L))\left(A_{v}(t)v^{2}+B_{v}(t)vv'+C_{v}(t)v'^{2}\right)\biggr\}.\label{J med fin}\end{align}
In this expression we use the definitions $\omega_{0}^{2}=\frac{4\Omega\eta}{M\pi}$,
$\gamma_{\pm}=\gamma(1\pm D[k_{0}L])$, $\omega_{t}^{2}=\omega_{0}^{2}D(k_{0}L)$
and the time dependent functions are \begin{equation}
A_{\chi}(t)=\frac{\eta}{\pi}\int_{0}^{t}d\tau\int_{0}^{t}dsK(\tau-s)e^{-\gamma_{+}t}\frac{\sinh\left[s 
\gamma_{+}/2\right]\sinh\left[\tau \gamma_{+}/2\right]}
{\sinh\left[t \gamma_{+}/2\right]^{2}}e^{\gamma_{+}(s+\tau)/2}\label{Ax(t)}\end{equation}
\begin{equation}
B_{\chi}(t)=\frac{2\eta}{\pi}\int_{0}^{t}d\tau\int_{0}^{t}dsK(\tau-s)e^{-\gamma_{+}
t/2}\frac{\sinh\left[\tau \gamma_{+}/2\right]\sinh\left[(t-s)
\gamma_{+}/2\right]}{\sinh\left[t\gamma_{+}/2\right]^{2}}e^{\gamma_{+}(s+\tau)/2}\label{Bx(t)}\end{equation}
\begin{equation}
C_{\chi}(t)=\frac{\eta}{\pi}\int_{0}^{t}d\tau\int_{0}^{t}dsK(\tau-s)
\frac{\sinh\left[(t-s)\gamma_{+}/2\right]\sinh\left[(t-\tau)
\gamma_{+}/2\right]}{\sinh\left[t
\gamma_{+}/2\right]^{2}}e^{\gamma_{+}(s+\tau)/2},\label{Cx(t)}\end{equation}
\begin{equation}
A_{v}(t)=\frac{\eta}{4\pi}\int_{0}^{t}d\tau\int_{0}^{t}dsK(\tau-s)e^{-\gamma_{-}t}\frac{\sinh\left[\frac{s}{2}
\sqrt{\gamma_{-}^{2}-4\omega_{t}^{2}}\right]\sinh\left[\frac{\tau}{2}
\sqrt{\gamma_{-}^{2}-4\omega_{t}^{2}}\right]}{\sinh\left[\frac{t}{2}\sqrt{\gamma_{-}^{2}-
4\omega_{t}^{2}}\right]^{2}}e^{\gamma_{-}(s+\tau)/2}\label{Av(t)}\end{equation}
\begin{equation}
B_{v}(t)=\frac{\eta}{2\pi}\int_{0}^{t}d\tau\int_{0}^{t}dsK(\tau-s)e^{-\gamma_{-}t/2}
\frac{\sinh\left[\frac{\tau}{2}\sqrt{\gamma_{-}^{2}-4\omega_{t}^{2}}\right]\sinh\left[\frac{(t-s)}{2}
\sqrt{\gamma_{-}^{2}-4\omega_{t}^{2}}\right]}{\sinh\left[\frac{t}{2}\sqrt{\gamma_{-}^{2}-
4\omega_{t}^{2}}\right]^{2}}e^{\gamma_{-}(s+\tau)/2}\label{Bv(t)}\end{equation}
\begin{equation}
C_{v}(t)=\frac{\eta}{4\pi}\int_{0}^{t}d\tau\int_{0}^{t}dsK(\tau-s)\frac{\sinh\left[\frac{(t-s)}{2}
\sqrt{\gamma_{-}^{2}-4\omega_{t}^{2}}\right]\sinh\left[\frac{(t-\tau)}{2}\sqrt{\gamma_{-}^{2}-
4\omega_{t}^{2}}\right]}{\sinh\left[\frac{t}{2}\sqrt{\gamma_{-}^{2}-4\omega_{t}^{2}}\right]^{2}}
e^{\gamma_{-}(s+\tau)/2}.\label{Cv(t)}\end{equation}
\end{widetext} $\mathcal{N}(t)$ is a time dependent coefficient
resulting from  the fluctuation around the classical path and will
be determined imposing the normalization condition to the final
reduced density operator. With the propagator (\ref{J med fin}) we
have all the tools needed for studying the time evolution of the
reduced density matrix which we do in the next section with a
particular example.

\section{Evolution of a Two Particle State}

Now, we want to study the consequences of the effective interaction
induced by the reservoir in the time evolution of a system formed
by two particles, without any direct interaction between them. Our
atention will be focused on the entanglement dynamics of a bipartite
system coupled to an oscillator bath and, for clarity, a brief introduction
about how to measure entanglement in Gaussian states is necessary.

A Gaussian state can be described using the characteristic function,
which in general can be written as \begin{equation}
\tilde{W}(\mathbf{X})=\exp\left\{ -\frac{1}{2}\mathbf{X}\boldsymbol{\Lambda}\mathbf{X}^{T}\right\} ,\label{caract geral}\end{equation}
where $\mathbf{X}=\left(\nu_{1},\lambda_{1},\nu_{2},\lambda_{2}\right)$
and $\boldsymbol{\Lambda}$ is the covariance matrix with elements
$\Lambda_{ij}=\left\langle X_{i}X_{j}+X_{j}X_{i}\right\rangle /2$.
The characteristic function is related to the density matrix through\emph{
}the transform\begin{equation}
\tilde{W}(\boldsymbol{\lambda},\boldsymbol{\nu})=\int_{-\infty}^{\infty}\exp\left(-i\boldsymbol{\lambda}\mathbf{x}/\hbar\right)\left\langle \mathbf{x}-\frac{\boldsymbol{\nu}}{2}\right|\hat{\rho}\left|\mathbf{x}+\frac{\boldsymbol{\nu}}{2}\right\rangle d^{2}x,\label{caract funct}\end{equation}
where $\boldsymbol{\lambda}=\left(\lambda_{1},\lambda_{2}\right)$
and $\boldsymbol{\nu}=\left(\nu_{1},\nu_{2}\right)$ are a sort of
momentum and position variables respectively. The covariance matrix
contains all the useful information of the Gaussian state and determines
it completely. The covariance matrix shall be conveniently written
in terms of $2\times2$ block submatrices as \begin{equation}
\boldsymbol{\Lambda}=\left(\begin{array}{cc}
A & C\\
C^{T} & B\end{array}\right).\label{CM blocks}\end{equation}
A very common quantifier of entaglement is the logarithmic negativity
$E_{\mathcal{N}}$ \cite{Adesso}, defined by

\begin{equation}
E_{\mathcal{N}}(\rho)=\max\left[0,-\ln2\tilde{\sigma}_{-}\right].\label{negat
log}\end{equation} where $\tilde{\sigma}_{\pm}$ are the symplectic
eigenvalues of the partial transposed density matrix that can be
written using the local symplectic invariants
$\textrm{det}\Lambda$, $\textrm{det}A$, $\textrm{det}B$,
$\textrm{det}C$ as \begin{equation}
\tilde{\sigma}_{\pm}=\frac{1}{\sqrt{2}}\left[\tilde{D}_{\Lambda}\pm\sqrt{\tilde{D}_{\Lambda}^{2}-4\det\Lambda}\right]^{\frac{1}{2}},\label{nu
til}\end{equation} where $\tilde{D}_{\Lambda}=\det A+\det B-2\det
C$. This expression quantifies directly the violation of the PPT
(\emph{positive partial transposition}), a necessary and
sufficient condition of separability.

We choose as initial state a Gaussian density matrix of two modes
caracterized by the squeeze parameter $z$. The initial density
matrix written in the variables $r$, $u$, $v$, $\chi$, is
\begin{align}
\tilde{\rho}&(r',\chi',u',v',0)=\nonumber\\
&\frac{1}{2\pi\sigma^{2}}\exp\left\{
-\frac{e^{-2z}}{4\sigma^{2}}\left(4r'^{2}+\chi'^{2}\right)-\frac{e^{2z}}{4\sigma^{2}}\left(u'^{2}+v'^{2}/4\right)\right\}
.\label{rho init new}\end{align} The density matrix at $t>0$ is
obtained from the expression (\ref{redu two P}), where the
propagator $J$ and the initial density matrix are given
respectively by (\ref{J med fin}) and (\ref{rho init new}), with the
suitable change of variables. The integrals indicated in
(\ref{redu two P}) can be solved exactly for all the limits
mentioned above since the involved variables appear only up to
second order. After the evolution of those integrals the
covariance matrix is easily computed by the Fourier transform
(\ref{caract funct}) and the results of (\ref{negat log}) are
presented for various parameter combinations in the figures below.

\begin{figure}
\begin{centering}
\includegraphics{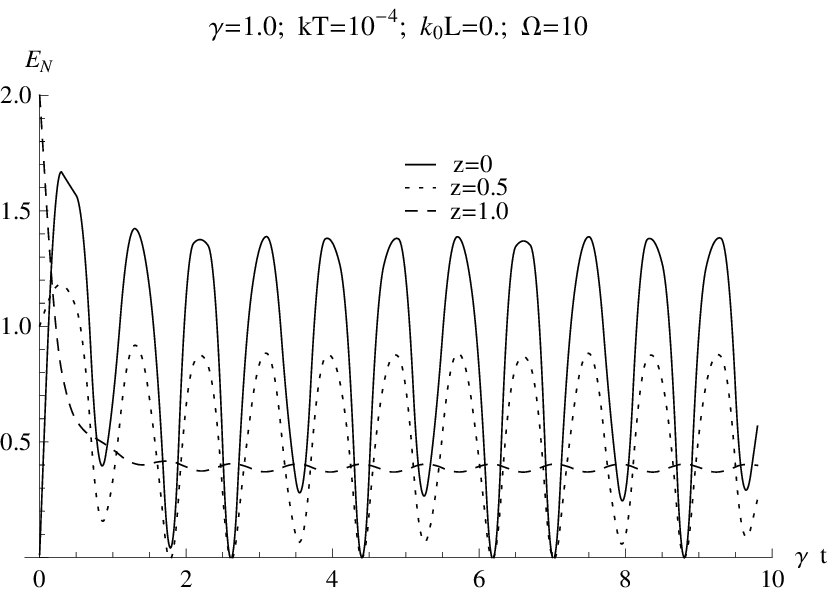}
\par\end{centering}

\caption{\label{fig:Enclose1}logarithmic negativity for close
particles as a function of time, the parameters are specified at
the top of the figure. }

\end{figure}

First, lets us observe the evolution of the logarithmic negativity
when the two parts of the system of interest are in the local
(short distance) regime. In figure (\ref{fig:Enclose1}) we have
the logarithmic negativity as a function of time for low
temperatures and three values of the squeeze parameter. The
dependence on the initial entanglement is evident in this figure.
For a initial separable system ($z=0$) the environment rapidly
induces entanglement and on the average maintains a  value higher
than that for systems initially entangled in which the decoherence
process drastically reduces  the logarithmic negativity. The
oscillatory behavior is due to the effective interaction potential
induced by the environment \cite{Duarte} and it is remarkable
that in all the cases plotted in figure (\ref{fig:Enclose1}) there
exist a remanent finite entanglement.

\begin{figure}
\begin{centering}
\includegraphics{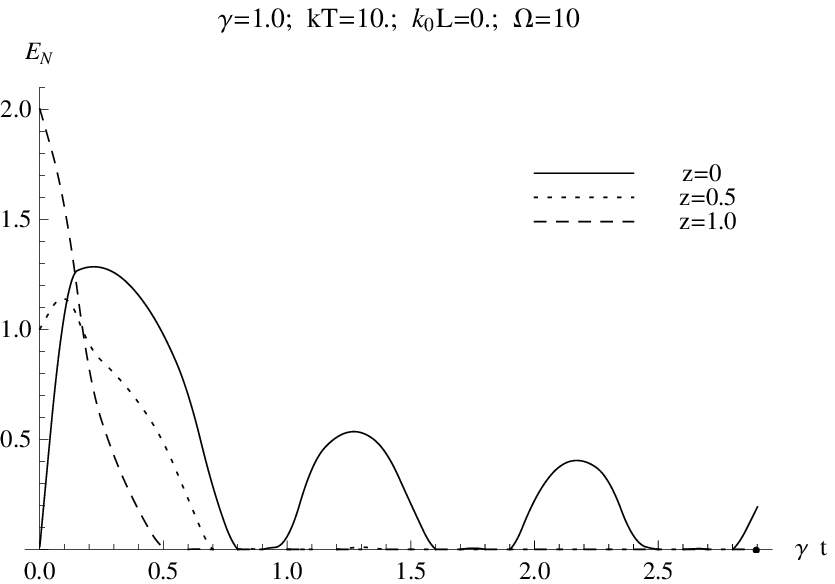}
\par\end{centering}

\caption{\label{fig:Enclose2}logarithmic negativity for close
particles as a function of time. The parameters are specified at
the top of the figure . }

\end{figure}

Figure (\ref{fig:Enclose2}) show us that the remanent entanglement
is an effect associated with low temperatures and when the
temperature is raised the initial entanglement always go to zero.
Again in this case, the initially separable state is the most
robust and maintains the induced entanglement for longer times.

\begin{figure}
\begin{centering}
\includegraphics{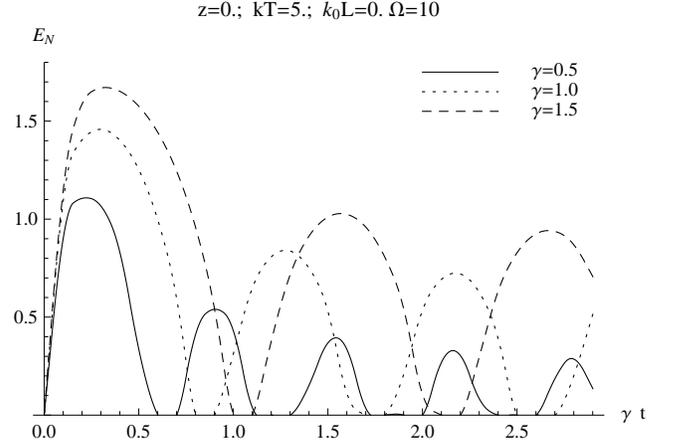}
\par\end{centering}

\caption{\label{fig:Enclose3}logarithmic negativity for close
particles and finite temperature as a function of time. The
parameters are specified at the top of the figure . }

\end{figure}

When the particles are very close together the most important influence
of the reservoir is the effective interaction induced between them
and the decoherence effect only becomes important when the temperature
rises. In figure (\ref{fig:Enclose3}) the logarithmic negativity
is plotted for three values of the dissipative constant $\gamma$.
The average amount of entanglement is higher for higher values of
$\gamma$, i.e. stronger interaction with the reservoir. Results similar
to this, but following other formalism and always considering a bilinear
system-reservoir coupling, have been presented by other autors \cite{Horhammer}.

\begin{figure}
\begin{centering}
\includegraphics{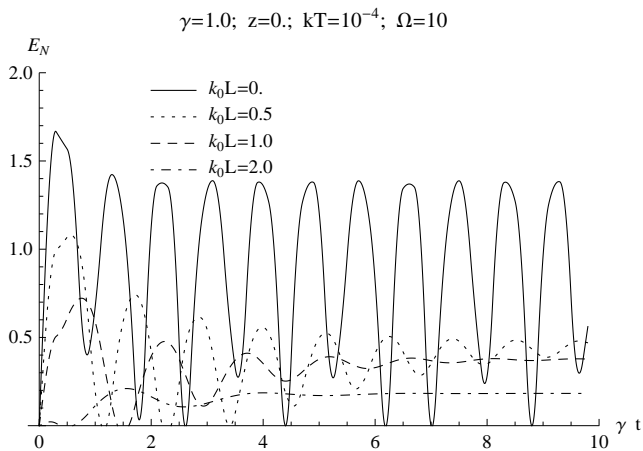}
\par\end{centering}

\caption{\label{fig: Enmed1}Influence of the distance between the
particles on the logarithmic negativity for low temperature. The
parameters are specified at the top of the figure. }

\end{figure}

Up to this point the results showed correspond to bipartite system
for which it is possible to define an average distance $L$ between
the two parts and this distance is very small when compared with
the characteristic length of the reservoir, $k_{0}^{-1}$. This
approximation assumes that the relative coordinate is completely
free from any dissipative influence. In a more realistic situation
a dissipative dynamics for the center of mass and relative motions
is expected and it is plausible to find a monotonically decreasing
remanent entanglement when the distance between the particles is
increased. In figure (\ref{fig: Enmed1}) we can observe the
influence of the distance between the particles on the evolution
of entanglement. It seems there is a competition between the
process of entanglement induction and decoherence. In particular,
when the distance increases the average entanglement is reduced
and the oscillations become damped. In the curve for $k_{0}L=2.0$
a delay in establishing  entanglement can be seen. Increasing the
distance $L$ reduces the induced interaction and increases the
delay time. In the limit $k_{0}L\to\infty$ the delay time goes to
infinity and for finite distances an asymptotically entangled
state when the temperature is low enough is observed.

\begin{figure}
\begin{centering}
\includegraphics{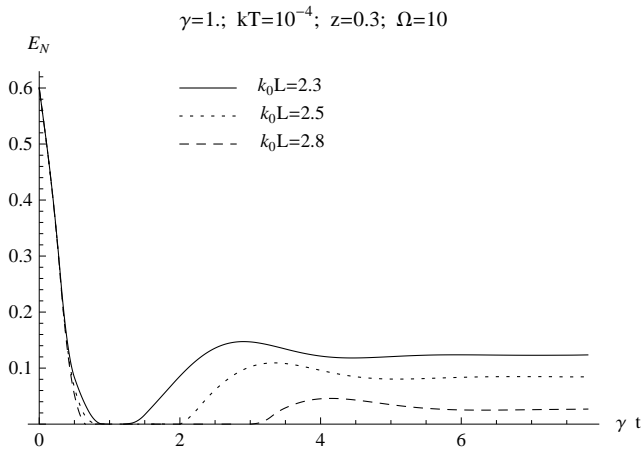}
\par\end{centering}

\caption{\label{fig: Enmed2}Influence of the distance on the revival of quantum
correlations }

\end{figure}

If the inital state is entangled the delay time is a revival time.
The initial entanglement is lost by decoherence due to the action
of the reservoir and after that the system is again entangled by
the induced interaction. This effect is only appreciable at low
temperatures. Then, it is possible to tune the induced
entanglement process varying the parameters and, in particular,
controlling the distance between the subsystems as shown in figure
(\ref{fig: Enmed2}).

\section{Conclusions}

In this paper we presented a system-plus-reservoir model with a
nonlinear coupling in the system coordinates which, in the
adequate limit, allows us to reproduce the phenomenological
results known for the quantum dynamics of a Brownian particle. The
model was extended for dealing with a system of two uncoupled
particles immersed in a common environment. The choice made for
the behavior of the bath response function, in order to reproduce
the dynamics of the quantum brownian motion for each particle when
isolated, naturally generates an effective interaction which
depends on the average distance between the particles and the
dissipation constant $\gamma$, as can be seen from the exact
density operator. When the particles are very close together the
interaction is basically harmonic and the amplitude of the induced
quantum correlations oscillates without decaying when the
temperature is low. In this approximation the dynamics is local
and the system preserves the induced entanglement. It is worth
noticing that when the initial state is entangled, the system
evolution is less sensitive to the induced correlations and more
sensitive to the decoherence effects. For interparticle distances
comparable with the characteristic length $k_{0}^{-1}$ of the
reservoir other phenomena appear, in particular we have the
possibility controlling the time interval between the total lack
of initial entanglement and the induced entanglement.

The results presented here may be relevant for a better
understanding of the bipartite systems with regard to the theory
of quantum information and the extension to deal with many
Brownian particles might contribute to the study of quantum
correlations of multipartite systems, and of the behavior of
complex many particles systems in general.

We kindly acknowledge Fundação de Amparo à Pesquisa do Estado de São
Paulo (FAPESP) and Conselho Nacional de Desenvolvimento Científico
e Tecnológico (CNPq), through the Millenium Institute for Quantum
Information, for the financial support.


\begin{thebibliography}{10}
\bibitem{Duarte}O. S. Duarte and A. O. Caldeira, Phys. Rev. Lett.
\textbf{97} 250601 (2006)

\bibitem{annals}A. O. Caldeira and A. J. Leggett, Ann. Phys. (N.Y.)
\textbf{149}, 374 (1983); \emph{ibid}. \textbf{153}, 445(E) (1983).

\bibitem{physica}A. O. Caldeira and A. J. Leggett, Physica (Amsterdam)\textbf{
121A}, 587 (1983).

\bibitem{Leggett}A. J. Leggett et. al., Rev. Mod. Phys. \textbf{59},1
(1987).

\bibitem{Feynman-Vernon}R. P. Feynman and A. F. L. Vernon, Ann. Phys.
(N.Y.) \textbf{24}, 118 (1963)

\bibitem{Fisher} M. P. A. Fisher and W. Zwerger, Phys. Rev. B \textbf{32},
6190 (1985).

\bibitem{Garg}A. Garg, J. N. Onuchic, and V. Ambegaokar , J. Chem.
Phys. \textbf{83}, 4491 (1985).

\bibitem{solitons} A.H. Castro Neto and A.O. Caldeira Phys. Rev.
E\textbf{48}, 4037 (1993).

\bibitem{Weiss}U. Weiss, Quantum Dissipative Systems, World Scientific
(1993)

\bibitem{Hedegard}P. Hedegard and A. O. Caldeira, Physica Scripta
\textbf{35}, 609 (1987).

\bibitem{Guinea}F. Guinea, Phys. Rev. Lett. \textbf{53}, 1268 (1984).

\bibitem{Ambegaokar}V. Hakim and V. Ambegaokar, Phys. Rev. A \textbf{32},
423 (1985).

\bibitem{Fetter}A. L. Fetter and J. D. Walecka, Quantum Theory of
Many Particles Systems, McGraw-Hill (1971)

\bibitem{Adesso}G. Adesso and F. Illuminati, Phys. Rev. A, \textbf{72},
032334 (2005)

\bibitem{Horhammer}Christian H\"orhammer and Helmut B\"uttner, Phys.
Rev. A, \textbf{77},\textbf{ }042305 (2008)

\bibitem{J Paz}Juan Pablo Paz and Augusto J. Roncaglia, Phys. Rev.
Lett. \textbf{100 }220401 (2008)
\end{thebibliography}
\end{document}